\documentclass{aastex63}

\received{XXX, 2020}
\revised{XXX, 2020}
\accepted{21 April 2020}
\submitjournal{ApJS}

\shorttitle{Physical association of $\gamma$-ray blazars}
\shortauthors{de Menezes et al.}

\graphicspath{{./}{figures/}}

\begin{document}

\title{On the physical association of \textit{Fermi}-LAT blazars with their low-energy counterparts}

\correspondingauthor{Raniere de Menezes}
\email{Contact e-mail: raniere.m.menezes@gmail.com}

\author[0000-0001-5489-4925]{Raniere de Menezes}
\affiliation{Universidade de S\~ao Paulo, Departamento de Astronomia,
Rua do Mat\~{a}o, 1226, S\~ao Paulo, SP 05508-090, Brazil}
\affiliation{Dipartimento di Fisica, Universit\`a degli Studi di Torino, via Pietro Giuria 1, I-10125 Torino, Italy}

\author{Raffaele D'Abrusco}
\affiliation{Center for Astrophysics | Harvard \& Smithsonian, 60 Garden Street, Cambridge, MA 20138, USA}

\author{Francesco Massaro}
\affiliation{Dipartimento di Fisica, Universit\`a degli Studi di Torino, via Pietro Giuria 1, I-10125 Torino, Italy}
\affiliation{Istituto Nazionale di Fisica Nucleare, Sezione di Torino, I-10125 Torino, Italy}
\affiliation{INAF-Osservatorio Astrofisico di Torino, via Osservatorio 20, 10025 Pino Torinese, Italy}
\affiliation{Consorzio Interuniversitario per la Fisica Spaziale (CIFS), via Pietro Giuria 1, I-10125, Torino, Italy}

\author{Dario Gasparrini}
\affiliation{Istituto Nazionale di Fisica Nucleare, Sezione di Roma ``Tor Vergata", I-00133 Roma, Italy}
\affiliation{Space Science Data Center - Agenzia Spaziale Italiana, Via del Politecnico, snc, I-00133, Roma, Italy}

\author{Rodrigo Nemmen}
\affiliation{Universidade de S\~ao Paulo, Departamento de Astronomia,
Rua do Mat\~{a}o, 1226, S\~ao Paulo, SP 05508-090, Brazil}

\begin{abstract}

Associating $\gamma$-ray sources to their low-energy counterparts is one of the major challenges of modern $\gamma$-ray astronomy. In the context of the Fourth \textit{Fermi} Large Area Telescope Source Catalog (4FGL), the associations rely mainly on parameters as apparent magnitude, integrated flux, and angular separation between the $\gamma$-ray source and its low-energy candidate counterpart. In this work we propose a new use of likelihood ratio and a complementary supervised learning technique to associate $\gamma$-ray blazars in 4FGL, based only on spectral parameters as $\gamma$-ray photon index, mid-infrared colors and radio-loudness. In the likelihood ratio approach, we crossmatch the WISE Blazar-Like Radio-Loud Sources catalog with 4FGL and compare the resulting candidate counterparts with the sources listed in the $\gamma$-ray blazar locus to compute an association probability for 1138 counterparts. In the supervised learning approach, we train a random forest algorithm with 869 high confidence blazar associations and 711 fake associations, and then compute an association probability for 1311 candidate counterparts. A list with all 4FGL blazar candidates of uncertain type associated by our method is provided to guide future optical spectroscopic follow up observations.

\end{abstract}


\section{Introduction}
\label{sec:intro}

The association of $\gamma$-ray sources with their low-energy counterparts is a long-standing challenge since the beginning of modern $\gamma$-ray astronomy \citep{fichtel1994first}. Its major underlying difficulty is related to the large positional uncertainty of $\gamma$-ray observations. Even in the era of the \textit{Fermi} Large Area Telescope \citep[LAT;][]{atwood2009LAT}, the positional uncertainty of $\gamma$-ray sources range from a couple of arcminutes up to $\sim 1^{\circ}$ \citep{lat20194fgl}. Then, according to the recent release of the \textit{Fermi}-LAT Fourth Source Catalog \citep[4FGL;][]{lat20194fgl}, $\sim 25\%$ of its sources lack an assigned low-energy counterpart and thus have uncertain nature. Although the fraction of unassociated $\gamma$-ray sources (UGSs) is still large, 4FGL presents a modest improvement in comparison with its previous releases, which had $\sim30\%$ of unassociated sources \citep{abdo2010_1FGL,nolan2012_2FGL,acero2015_3FGL}. The largest population of associated $\gamma$-ray sources is dominated by pulsars, pulsar wind nebulae and supernova remnant in the Galactic plane, and by blazars in the extragalactic sky \citep{lat20194fgl}

Several studies searching for the counterparts of $\gamma$-ray sources have been performed in the past decade \citep{abdo2010_1FGL,nolan2012_2FGL,acero2015_3FGL}. Dedicated follow up observations at radio, infrared and X-rays allowed for the identification of potential counterparts for several \textit{Fermi}-LAT sources, for which optical spectra were then collected to establish their nature \citep{massaro2015refining,massaro2016gamma,crespo2016opticalCampVI,pena2017optical,marchesini2019optical,pena2019optical,deMenezes2020_OptCampX}. Once a candidate counterpart is found, it can be considered in the association methods of the \textit{Fermi}-LAT catalogs. All association methods used by the \textit{Fermi}-LAT Collaboration \citep{lat20194fgl} consist in computing the association probability (AP) based mainly on the angular separation between the center of the $\gamma$-ray source and the position of its candidate counterpart, thus being mostly a geometrical approach and neglecting physical properties of candidate counterparts, such as colors, spectral shape and radio-loudness, to name a few. When a candidate counterpart has AP $> 80\%$ in 4FGL, the source is considered associated.

In this work we propose two different association procedures, both independent of angular separation and relying mainly on the $\gamma$-ray-mid-infrared connection of candidate blazar counterparts selected from the WISE Blazar-Like Radio-Loud Sources catalog \citep[WIBRaLS;][]{dAbrusco2019WIBRaLS}. WIBRaLS is a catalog of radio-loud candidate blazars whose WISE mid-infrared (MIR) colors are selected to be consistent with MIR colors of confirmed $\gamma$-ray emitting blazars \citep{massaro2011blazars_IR,dabrusco2012infrared}. We compute the APs with two different methods: i) the likelihood ratio \citep[LR;][]{sutherland1992likelihood}, already adopted by the \textit{Fermi}-LAT Collaboration, but with a different setup; and ii) a random forest algorithm \citep[RF;][]{breiman2001random}. By assigning an AP to each candidate counterpart, we can schedule/program future optical spectroscopic follow-ups on these results, prioritizing those targets with higher AP.

Computing an AP for each counterpart candidate is crucial because all methods used to select candidate
blazars are statistical in nature and do not take into account the specific $\gamma$-ray properties of each \textit{Fermi} source to be associated but only collective features of a population of sources (in our case, the $\gamma$-ray emitting confirmed blazars in the blazar locus. See \S \ref{SampleSelectionLR}). Thus, association methods bridge the gap between collective behavior and each specific case, giving an intra-source prioritization -- should there be multiple candidates for the same $\gamma$-ray source -- and an inter-source prioritization to maximize the 
effectiveness of spectroscopic follow-ups.

Machine learning techniques have been used before in the context of \textit{Fermi}-LAT catalogs to i) predict the spectral class of UGSs \citep{doert2014search,parkinson2016classification,lefaucheur2017research,salvetti20173fglzoo} based only on the $\gamma$-ray properties available in the Second and Third \textit{Fermi} Source Catalogs \citep[2FGL and 3FGL, respectively;][]{nolan2012_2FGL,acero2015_3FGL}; ii) predict the nature of blazar candidates of uncertain type \citep[BCUs;][]{hassan2012gamma,chiaro2016blazar,kovavcevic2019optimizing}; iii) and even for spotting candidates of dark matter Galactic subhalos \citep{mirabal2012fermi}. This is the first time, however, that machine learning is used to associate \textit{Fermi}-LAT sources with their low-energy counterparts.

The idea underlying this work is that $\gamma$-ray blazars have some specific radio-MIR characteristics which can tell them between, e.g., two counterpart candidates lying in the same elliptical uncertainty region of a $\gamma$-ray source. Indeed, some sources associated in 4FGL have a very high AP with counterparts that are not the closest ones if compared with sources listed in the latest version of WIBRaLS \citep{dAbrusco2019WIBRaLS}. This can happen simple because the latest version of WIBRaLS was not taken into account when associating 4FGL sources, but highlights a possible bias in a method that associate sources based mainly on angular separation: if the real $\gamma$-ray source counterpart is not listed in one of the catalogs used by the association algorithms, then the method will simply choose the closest candidate.

The paper is organized as follows. In \S \ref{SampleSelectionLR} and \S \ref{SampleSelectionRF} we describe the samples used to carry out our analysis, providing basic details on both WIBRaLS and 4FGL. In \S \ref{sec:methodsLR} and \S \ref{sec:methods_RF} we describe the adopted methods followed by the achieved results in \S \ref{Results_LR} and \S \ref{Results_RF}. We conclude and summarize our work in \S \ref{discussion}. The WISE magnitudes are in the Vega system and are not corrected for the Galactic extinction, since, as shown in \cite{dAbrusco2014wibrals1}, such correction only affects the magnitude at 3.4 $\mu$m for sources lying close to the Galactic plane and it ranges between 2\% and 5\% of the magnitude, thus not affecting significantly the results. WISE bands are indicated as W1, W1, W3 and W4, and correspond respectively to the nominal wavelengths at 3.4, 4.6, 12, and 22 $\mu$m, while the colors are defined as $c_{12} = W1-W2$, $c_{23} = W2-W3$ and $c_{34} = W3-W4$. The adopted MIR radio-loudness parameter is defined as $q_{22} = \log (S_{22\mu m}/S_{radio})$, where $S_{22\mu m}$ is the flux density in the WISE W4 band, and $S_{radio}$ is the radio flux density at 1.4 GHz or at 843 MHz, depending on the radio survey in which the WIBRaLS radio counterpart is identified \citep[see][for more details]{dAbrusco2019WIBRaLS}.

\section{Samples used for the likelihood ratio method}
\label{SampleSelectionLR}

We apply the LR method to all 1311 sources listed in 4FGL with at least one counterpart in WIBRaLS, comparing their MIR colors and radio-loudness with those of the $\gamma$-ray blazars lying within the blazar locus \citep{dabrusco2013_blazarlocus}. 

The blazar locus is defined by a sample of confirmed \textit{Fermi}-LAT blazars listed in Roma-BZCAT \citep{massaro20155_BZCAT} and associated with WISE counterparts detected in all four filters W1, W2, W3 and W4. The locus is modeled in a three-dimensional space generated by the principal components of the MIR color-color-color distribution and describes the typical MIR colors of $\gamma$-ray blazars, thus being ideal for our purposes \citep[see, e.g.,][]{dabrusco2013_blazarlocus,dAbrusco2019WIBRaLS}.

\section{Samples used for the supervised learning method}
\label{SampleSelectionRF}

The training set of supervised learning algorithms needs to be representative of the expected distribution of outcomes in the test set, or the sample on which the trained algorithm will be applied. In this case, we train a RF algorithm by gathering high-confidence associations and not-associated (fake) counterparts. We do it by selecting all associated sources in 4FGL having a counterpart in WIBRaLS and with $\gamma$-ray statistical significance above $10\sigma$. Such high confidence $\gamma$-ray detections tend to have smaller error ellipses, and for this reason they are easier to correctly associate with low-energy counterparts. The final number of high confidence associations in the training set was 869. The fake associations were selected as the WIBRaLS sources not associated in 4FGL but lying within $0.5^{\circ}$ from 4FGL sources that are already associated with a low-energy counterpart, resulting in a total of 711 fake counterparts for the training sample. It is possible that a few of these fake associations are the real counterparts of the $\gamma$-ray sources.

\begin{figure*}
    \centering
    \includegraphics[width=\linewidth]{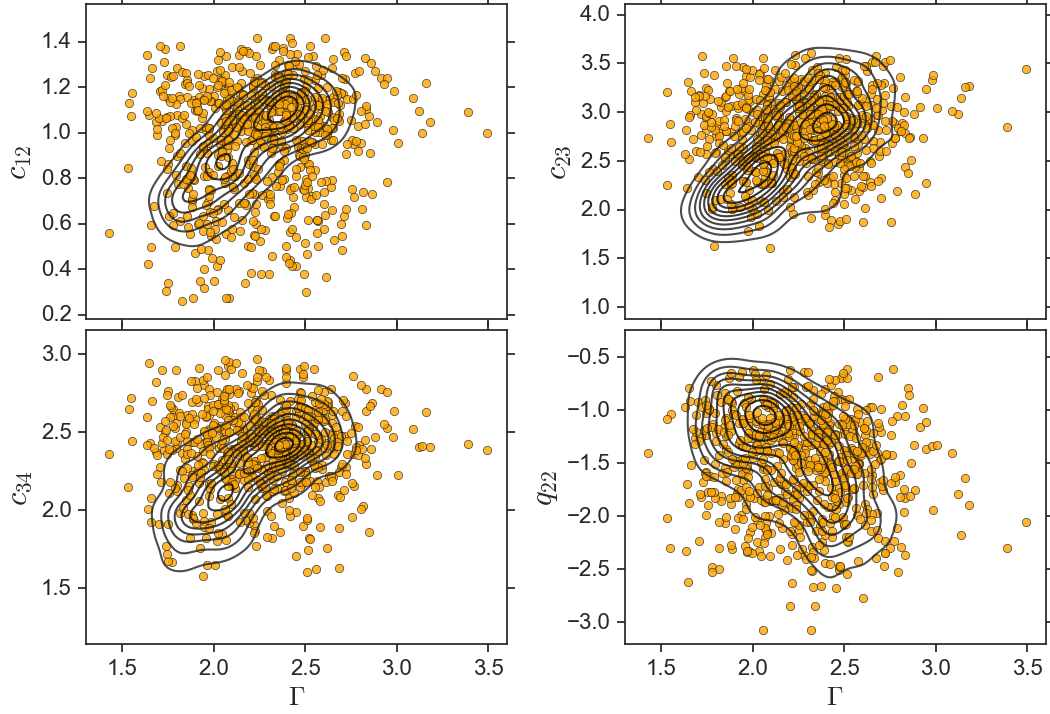}
    \caption{Training sample for the RF algorithm. The contours represent the distribution of the 869 high confidence associations, while the orange dots represent the fake associations. The spectral parameters for the training sample are shown in terms of the $\gamma$-ray photon index, $\Gamma$, the three MIR colors $c_{12}$, $c_{23}$ and $c_{34}$ (as defined in the last paragraph of \S \ref{sec:intro}) and the MIR radio-loudness $q_{22}$.}
    \label{fig:training_sample}
\end{figure*}

The panels in Figure \ref{fig:training_sample} show the differences between high confidence associations (contourns) and fake associations (orange dots) in the training sample. Based on these $\gamma$-ray-MIR characteristics, supervised learning algorithms can be trained to tell how likely a WIBRaLS source is associated as the counterpart of a $\gamma$-ray source. Once trained, we applied the RF method to a test sample made of the 1311 4FGL-WIBRaLS crossmatches, the same sample used in the LR method. 




\section{The likelihood ratio method}
\label{sec:methodsLR}

The LR technique was first used in the context of source association by \cite{richter1975search} and has been introduced and developed to look for possible counterparts among faint radio, infrared and X-ray sources \citep{wolstencroft1986identification,sutherland1992likelihood,masci2001new}. In the context of \textit{Fermi}-LAT, the LR between a candidate counterpart $i$ within the error ellipse of a 4FGL source $j$ is computed as \citep{ackermann2011_2LAC,Fermi_LAT_2019_4LAC}: 
\begin{equation}
    LR_{ij} = \frac{e^{-r^2_{ij}/2}}{NA} ,
    \label{eq:1}
\end{equation}
where $r_{ij} = \theta/(\sigma_{i}^2 + \sigma_{j}^2)^{1/2}$ is the normalized angular separation between the $\gamma$-ray source $j$ and candidate counterpart $i$, with $\theta$ being the angular separation between the counterpart and the center of the $\gamma$-ray source, $\sigma_i$ being the low-energy positional uncertainty and $\sigma_j \equiv \sigma_{95\%}/2.4477$ being the normalized geometric mean of the semi-major and semi-minor axes of the 95\% 4FGL confidence error ellipse. $N$ is the surface density of objects brighter than the candidate $i$ and $A$ is the solid angle encompassed by the 95\% confidence LAT error ellipse. Below we modify this method by taking into account only spectral properties of the candidate counterparts, as MIR colors and radio-loudness, and comparing these properties with what is expected for $\gamma$-ray blazars lying in the blazar locus \citep{dabrusco2013_blazarlocus}. In \S \ref{subsec:ang_sep} we add a dependence on the angular separation to this method to measure the impact of angular separation in our results.


The LR method we adopt to estimate the AP is a modification of the LR method described in \cite{sutherland1992likelihood} and \cite{ackermann2011_2LAC}. The adopted steps are as follows.

\begin{figure*}
    \centering
    \includegraphics[width=\linewidth]{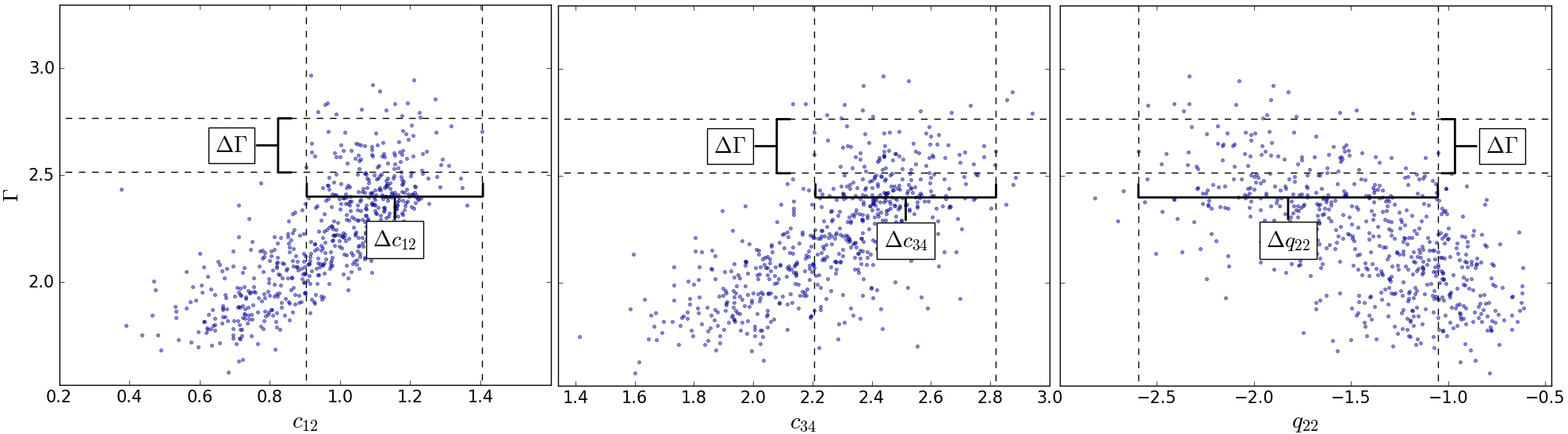}
    \caption{Region selected in the blazar locus for a source with $\gamma$-ray photon index $\Gamma = 2.65 \pm 0.10$. Given an interval $\Delta\Gamma$, only the sources within the ranges $\Delta c_{12}$, $\Delta c_{34}$ and $\Delta q_{22}$ can be selected as possible counterparts of $\gamma$-ray sources.}
    \label{fig:color_correlations}
\end{figure*}

\begin{figure}
    \centering
    \includegraphics[scale=0.5]{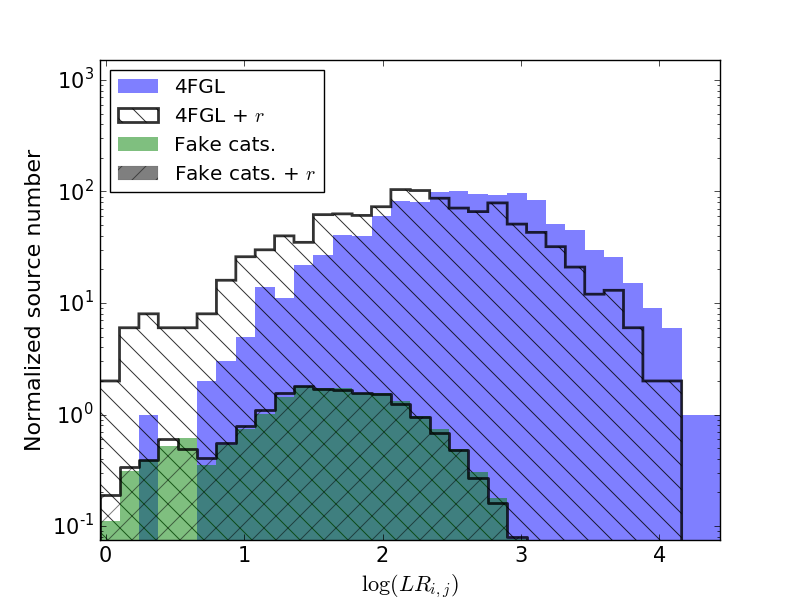}
    \caption{Distribution of the $\log (LR_{i,j})$ for the crossmatches between WIBRaLS and 4FGL ($G_{\mathrm{real}}$), in blue; and average distribution for the crosmatches between WIBRaLS and the fake $\gamma$-ray catalogs ($<G_{\mathrm{fake}}>$), in green. The hatched histograms represent the same distributions when taking angular separation, $r$, into account with the exponential behaviour shown in Equation \ref{eq:1}.}
    \label{fig:histos}
\end{figure}

\begin{figure}
    \centering
    \includegraphics[scale=0.5]{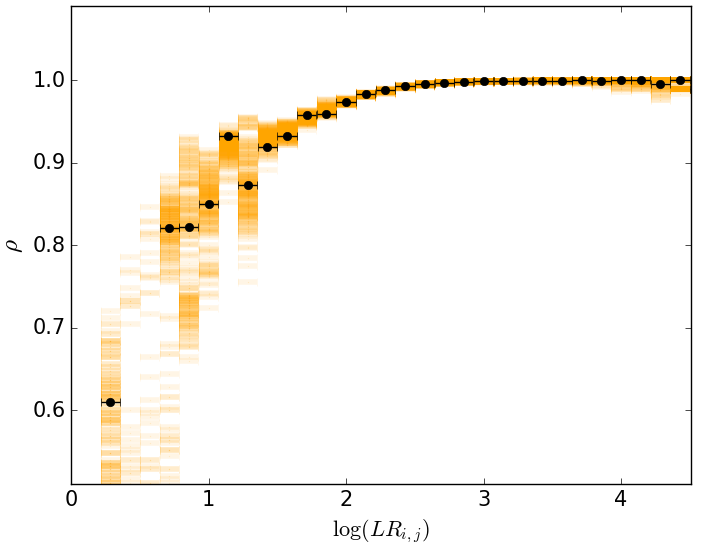}
    \caption{The reliability $\rho$ derived from the LR method as function of $\log(LR_{i,j})$. The black dots are the points adopted in this work, while the orange shadow represents the uncertainty region due to different choices of radii when computing the local surface density N($\Delta c_{12},\Delta c_{34},\Delta q_{22}$). In the region $\log(LR_{i,j}) > 1.5$, the impact of choosing different radii is negligible.}
    \label{fig:rho}
\end{figure}

\begin{enumerate}
    \item We crossmatch WIBRaLS with 4FGL to create the list of all $\gamma$-ray blazar candidates that lie within the positional uncertainty region at 95\% level of confidence of each \textit{Fermi}-LAT object. With a total of 1311 potential counterparts, the result of this crossmatch correspond to all of the 4FGL candidate counterparts listed in WIBRaLS.
    
    \item For each 4FGL source, we consider the $\gamma$-ray photon index interval $\Delta\Gamma = [\Gamma - \sigma_{\Gamma}, \Gamma + \sigma_{\Gamma}]$, where $\sigma_{\Gamma}$ is the uncertainty on the photon index as listed in 4FGL. Then, given the correlations between $\Gamma$, the MIR colors $c_{12}$ and $c_{34}$, and the MIR radio-loudness $q_{22}$ for the blazars in the $\gamma$-ray blazar locus \citep{dabrusco2012infrared}, we select from WIBRaLS only those sources with parameters within the blazar locus ranges $\Delta c_{12} = [c^{min}_{12}, c^{max}_{12}]$, $\Delta c_{34} = [c^{min}_{34} , c^{max}_{34}]$ and $\Delta q_{22} = [q^{min}_{22} , q^{max}_{22}]$ corresponding to the $\Delta\Gamma$ interval, as shown in Figure \ref{fig:color_correlations}. This step reduces the total number of candidate counterparts from 1311 to 1138.
    
    \item For each MIR counterpart $i$ of a 4FGL source $j$ we compute the LR using the following equation: 
    $$ \log LR_{i,j} = \log \left[ \frac{1}{2\pi\sigma_1\sigma_2}  \frac{Q(\Delta c_{12},\Delta c_{34},\Delta q_{22})}{N(\Delta c_{12},\Delta c_{34},\Delta q_{22})}   \right], $$
    where Q($\Delta c_{12},\Delta c_{34},\Delta q_{22}$) is the probability to find a WIBRaLS source with MIR colors and radio-loudness within the ranges $\Delta c_{12}$, $\Delta c_{34}$ and $\Delta q_{22}$ over the entire WIBRaLS catalog (i.e., it is the ratio between the total number of sources lying within the specified intervals and the total number of WIBRaLS sources), and N($\Delta c_{12},\Delta c_{34},\Delta q_{22}$) is the local surface density of WIBRaLS sources within a circle of $8^{\circ}$ radius centered in the 4FGL source and having MIR colors and radio-loudness within the specified ranges (i.e., it is the surface density of the background objects at the appropriate Galactic latitude). The positional uncertainty errors of the 95\% confidence regions of 4FGL sources are given by $\sigma_1$ and $\sigma_2$. Angular separation between the $\gamma$-ray source and its counterpart is not taken into account.
    
    \item To compute the AP based on the LR calculated above, we first generate 500 fake $\gamma$-ray catalogs by shifting the sky positions of each 4FGL source by a random value between $0.3^{\circ}$ and $3^{\circ}$ in a random direction of the sky. These small shifts in R.A. and Dec. guarantee that we preserve the inhomogeneity of the $\gamma$-ray sky, which has more sources concentrated towards the Galactic plane. We then repeat steps 2 and 3 above for all matches between the fake sources (belonging to the 500 generated catalogs) and WIBRaLS sources to compute the average distribution function ($<G_{\mathrm{fake}}(LR_{i,j})>$) of fake $\log LR_{i,j}$. The $\log LR_{i,j}$ distributions for the real and fake matches are shown in blue and green, respectively, in Figure \ref{fig:histos}. As a matter of comparison, we overplot the $\log LR_{i,j}$ distributions when adding the exponential dependence of angular separation, $r$, shown in Equation \ref{eq:1}, to our method (hatched distributions).
    
    \item We then compare the real and the fake distributions of $\log LR_{i,j}$ to determine the reliability for the real associations by computing: $$ \rho(LR_{i,j}) = 1 - \frac{<G_{\mathrm{fake}}(LR_{i,j})>}{G_{\mathrm{real}}(LR_{i,j})},   $$
    where $<G_{\mathrm{fake}}(LR_{i,j})>$ is the average $\log LR_{i,j}$ distribution for fake catalogs and $G_{\mathrm{real}}(LR_{i,j})$ is the $\log LR_{i,j}$ distribution for the real $\gamma$-ray source catalog, as shown in Figure \ref{fig:histos}. The reliability, $\rho$, computed according to this equation represents an approximate measurement of the AP for a potential counterpart having a given $\log LR_{i,j}$. In Figure \ref{fig:rho} we show the dependence of $\rho$ on $\log(LR_{i,j})$. Some fluctuations are observed for $\log(LR_{i,j}) < 1.5$, but the overall behaviour of the curve is clear: for $\log(LR_{i,j}) > 1.5$, basically all counterpart candidates have very high ($>95\%$) APs. The orange shadow in Figure \ref{fig:rho} represent the uncertainty region caused by choosing different radii when computing the local surface density (see step 3 above). This region was generated by ranging the radii from $3.5^{\circ}$ up to $15^{\circ}$ in 200 linearly spaced intervals. We observe that, in the region defined by $\log(LR_{i,j}) > 1.5$, the APs do not really depend on how we choose the background region radius.

\end{enumerate}

The APs obtained with this method range from 60\% to 100\%. Among the sources listed as identified in 4FGL--those for which the association to the low-energy counterpart is guaranteed--the APs are always above 99\%.

\section{The random forest method}
\label{sec:methods_RF}

In the context of supervised learning algorithms, we use a RF \citep{breiman2001random} to compute the AP of 4FGL counterparts. The RF method is an ensemble classifier that uses decision trees as building blocks for classification \citep{james2013introduction}. For classifying a new object, each tree in the forest chooses one class and, by aggregating the predictions of all decision trees, the RF makes a final prediction based on the choice made by the majority of the trees, thus improving the predictive capability and reducing the tendency of standard decision trees to overfit the training sample.

We train and estimate the accuracy of the RF method with the sample described in \S \ref{SampleSelectionRF} using cross-validation: the RF algorithm has been trained with 10 subsets, each containing 90\% of the training sample (where the sources are randomly chosen), and testing it on the complementary 10 subsets containing 10\% of the training sample, in a way that the final accuracy of $78.2\% \pm 3.3\%$ is simply the average ratio of correctly predicted observations to the total observations. The following attributes listed in 4FGL and WIBRaLS were used during the learning process: $\gamma$-ray power law photon index $\Gamma$, WISE colors $c_{12}$, $c_{23}$, and $c_{34}$, and radio-loudness $q_{22}$ computed with respect to the infrared flux in band W4. We tested several other parameters available in 4FGL and WIBRaLS, as $\gamma$-ray variability index and ratios between $\gamma$-ray flux densities (power law and logparabola spectral models) and radio flux density, but all of them presented a negligible impact on the results, with improvements in the algorithm accuracy of $\sim 3\%$ in the best case (i.e., $\gamma$-ray variability index), and sometimes only adding noise. Among the used parameters, the one with largest impact in the accuracy when combined with $\Gamma$ is $c_{12}$, followed by $c_{34}$, $q_{22}$ and $c_{23}$, respectively. As discussed in \S \ref{sec:intro}, we are neglecting angular separation between the center of the $\gamma$-ray sources and the position of the MIR sources listed in WIBRaLS, as we are proposing an association method that relies only on spectral properties.

We use the RF classifier available in the Python library \texttt{sklearn} \citep{scikit-learn}. The code fits several decision trees on sub-samples of the dataset and average them to improve the predicted accuracy and prevent over-fitting. To guarantee the stability of our results, we use 5000 trees and let the nodes grow until all leaves contain less than the minimum number of samples required to split an internal node. The APs obtained with the RF algorithm range from 5\% to 100\% and the lowest AP obtained for blazars listed as identified in 4FGL (all of them included in the training sample) is 70\%.

\section{Results from the likelihood ratio approach}
\label{Results_LR}

The total number of associations with AP $> 99\%$ (i.e., as high as the AP obtained for the identified 4FGL sources. See \S \ref{sec:methodsLR}) is 743 out of the 1138 original 4FGL-WIBRaLS crossmatches within the intervals $\Delta c_{12},\Delta c_{34}$ and $\Delta q_{22}$. This number increases to 1051 when considering counterparts with AP $> 95\%$. Among all the 1138 sources for which we compute an AP, 283 are associated in 4FGL as BCUs, 473 as BL Lacs, 350 as flat spectrum radio quasars (FSRQs) and the rest of them are divided into a few non-blazar extragalactic and unknown objects. We found 5 UGSs with a counterpart in WIBRaLS and with APs ranging from 93\% up to 99\%. The BCUs and UGSs associated here--specially those with APs as high as the identified 4FGL sources--are promising targets for future optical spectroscopic follow up missions. In fact, optical spectroscopic observations for some of them are available in \cite{deMenezes2020_OptCampX} and Pe\~{n}a-Herazo et al. (2020, in prep.), where all of the sources have blazar-like optical spectra. All BCUs and UGSs associated here are listed in Tables \ref{tab:BCUs} and \ref{tab:UGSs} in Appendix \ref{ap:tables}.

Among the $\gamma$-ray sources associated here (with any AP), 22 of them have different counterparts in 4FGL which are not listed in WIBRaLS (see Table \ref{tab:double_associations}). Probably due to the exponential dependence with angular separation (see Eq. \ref{eq:1}) adopted in 4FGL, the 22 counterparts listed in 4FGL are generally closer to the center of the $\gamma$-ray sources than the WIBRaLS counterparts selected based on MIR colors. Furthermore, 165 of the 1138 WIBRaLS counterpart candidates are not associated by the LR method adopted in 4FGL. These sources were associated in 4FGL mainly via the Bayesian method \citep{lat20194fgl}. Additionally, no correlation is observed between the APs computed here and the APs computed with the LR method adopted in 4FGL, indicating that the methods are completely independent.

\begin{table}
    \centering
    \begin{tabular}{|l|l|l|r|r|}
\hline
  \multicolumn{1}{|c|}{4FGL name} &
  \multicolumn{1}{c|}{Association WIBRaLS} &
  \multicolumn{1}{c|}{Association 4FGL} &
  \multicolumn{1}{c|}{AP LR WIB} &
  \multicolumn{1}{c|}{AP LR 4FGL} \\
\hline
   J0119.9+4053 & J011947.90+405418.4 & CRATES J012018+405314 & 0.97 & --\\
   J0211.1-0646 & J021116.95-064419.9 & LEDA 1029376 & 0.99 & 0.96\\
   J0237.7+0206 & J023737.97+020742.5 & PKS 0235+017 & 0.93 & --\\
   J0402.9+6433 & J040254.43+643510.0 & 1RXS J040301.8+643446 & 0.99 & 0.83\\
   J0640.9-5204 & J064111.26-520232.2 & ERC217 G261.25-22.62 & 0.99 & --\\
   J0725.6-3530 & J072548.16-353041.8 & NVSS J072547-353039 & 0.99 & --\\
   J0801.3-0617 & J080141.07-060535.2 & CRATES J080140-054037 & 0.61 & --\\
   J0807.7-1206 & J080730.69-120608.9 & CRATES J080736.06-120745.9 & 0.93 & 0.92\\
   J1122.0-0231 & J112213.71-022914.0 & 2QZ J112156-0229 & 0.96 & --\\
   J1136.3-0501 & J113547.40-050804.8 & NVSS J113607-050156 & 0.86 & --\\
   J1145.7+0453 & J114631.75+045819.2 & PKS 1142+052 & 0.93 & 0.89\\
   J1300.4+1416 & J130020.92+141718.4 & OW 197 & 0.96 & 0.95\\
   J1305.3+5118 & J130645.87+511655.0 & IERS B1303+515 & 0.83 & --\\
   J1516.8+2918 & J151641.71+291816.5 & RGB J1516+293 & 0.97 & 0.89\\
   J1518.6+0614 & J151847.70+061259.0 & TXS 1516+064 & 0.96 & 0.96\\
   J1550.8-1750 & J155053.18-175618.4 & TXS 1548-177 & 0.97 & 0.88\\
   J1734.0+0805 & J173510.44+080831.0 & 2MASS J17340287+0805237 & 0.86 & 0.83\\
   J1807.1+2822 & J180712.91+282059.5 & WISE J180634.08+281908.0 & 0.86 & --\\
   J1953.0-7025 & J195306.71-702428.7 & PKS 1947-705 & 0.99 & 0.96\\
   J2110.2-1021 & J211038.30-103337.2 & PKS 2107-105 & 0.86 & 0.95\\
   J2209.8-5028 & J221040.80-502652.5 & PMN J2210-5030 & 0.96 & --\\
   J2329.7-2118 & J232940.19-211345.0 & PKS 2327-215 & 0.97 & 0.88\\
\hline\end{tabular}
    \caption{List of sources with associations provided with the LR method developed here and having a different association in 4FGL. As expected, most (but not all) of the associations listed in 4FGL are closer to the $\gamma$-ray emission center than the WIBRaLS candidate counterparts selected based on MIR colors. Starting from the left side, the columns are i) the name of the $\gamma$-ray source in 4FGL, ii) the WISE name of the WIBRaLS candidate counterpart selected based on MIR colors, iii) the candidate counterpart listed in 4FGL, iv) the AP obtained with the LR method developed here, and v) the AP obtained with the LR used in 4FGL.}
    \label{tab:double_associations}
\end{table}

\subsection{Considering angular separation}
\label{subsec:ang_sep}

When we consider angular separation (with an exponential dependence as shown in Equation \ref{eq:1}) together with WISE MIR colors and $q_{22}$, the LR distributions look like the hatched histograms in Figure \ref{fig:histos}. As the distributions are very similar, the APs resulting when considering angular separation are also similar. The total number of associated counterparts in this case is 580 with AP $> 99\%$, or 1071 when considering AP $> 95\%$. Furthermore, all sources with AP $< 95\%$ have low $\gamma$-ray statistical significance ($<10\sigma$), which generally implies in large positional and $\gamma$-ray photon index errors.

\begin{figure}
    \centering
    \includegraphics[scale=0.55]{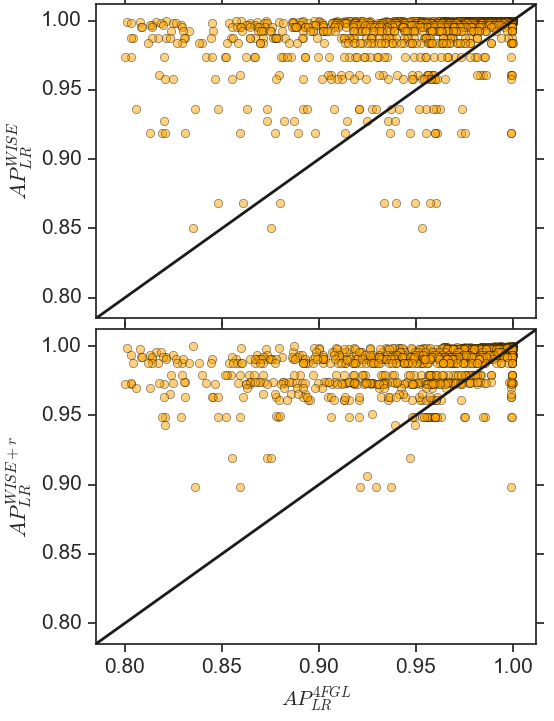}
    \caption{Comparison between the APs obtained with our LR method neglecting (upper panel) and considering (bottom panel) angular separation and the APs obtained with the LR method used in 4FGL. The APs computed by our method tend to be higher than those listed in 4FGL paper.}
    \label{fig:LR_comparison}
\end{figure}

In the upper panel of Figure \ref{fig:LR_comparison} we compare the APs obtained with the LR method developed here with those obtained with the LR adopted in 4FGL. In the bottom panel we show the same comparison, but when taking into account the angular separation, $r$, in the LR method. In both cases our method tend to give higher APs.

\section{Results from the random forest approach}
\label{Results_RF}

We found a total of 960 associations with AP $> 70\%$ (i.e., in the same range of the identified 4FGL sources. See \S \ref{sec:methods_RF}) out of 1311 test sources (see \S \ref{SampleSelectionRF}). Furthermore, the number of counterparts with AP $> 50\%$ increases to 1109. Due to the low accuracy presented by the RF method (see \S \ref{sec:methods_RF}), we use these results only as a comparison to the LR method described in the previous section. The APs for the BCUs and UGSs associated with the RF are shown in Tables \ref{tab:BCUs} and \ref{tab:UGSs} in Appendix \ref{ap:tables}.

When comparing the APs obtained with the RF algorithm and those obtained with the LR methods used here and in 4FGL, we found no significant correlation, although the APs computed with the RF approach tend to be smaller and distributed over a larger range. The major difference between both methods is the way we feed the two algorithms with the data. In Figure \ref{fig:training_sample} it is clear that several fake associations (orange dots) occupy the same region as the high confidence associations, thus counterparts lying within the high confidence association region (black contours) still have a high chance of being a fake association in the RF approach. The same does not happen with the LR method. Furthermore, based on how we defined the training set and the fake associations, we are solving an intrinsically more complicated task with the RF approach than with the LR method.

\section{Discussion and conclusions}
\label{discussion}

Associating $\gamma$-ray sources with their low-energy counterparts is challenging. Given the absence of multiwavelength monitoring and optical spectroscopic data within the error ellipses of many 4FGL sources, we have to recur to statistical methods when associating $\gamma$-ray sources to their counterparts. 

In this work we propose an alternative to the LR association method adopted by the \textit{Fermi}-LAT Collaboration \citep{lat20194fgl}. Here, the APs are independent of angular separation and apparent magnitude, and rely only on spectral parameters such as MIR colors and radio-loudness. As a complementary association method, we trained a RF algorithm to associate $\gamma$-ray sources to their counterparts. Both methods, however, are suited only for associating blazar candidates to $\gamma$-ray sources. Other types of $\gamma$-ray sources, like pulsars and starburst galaxies (to name a few), can not be directly associated with the proposed methods.

In the LR method adopted here we naturally loose some true blazar associations which have MIR colors just outside the blazar locus (i.e., the region of the WISE color-color-color diagram populated by confirmed $\gamma$-ray blazars). The sources associated here, however, are very good $\gamma$-ray blazar candidates and excellent targets for future optical spectroscopic follow ups. Furthermore, as all sources in WIBRaLS are radio-loud with respect to the $q_{22}$ parameter \citep{dAbrusco2019WIBRaLS}, it is likely to find many BL Lac-galaxy dominated among the sources excluded by our method. It is really unlikely that these sources have no sign of nuclear activity, as the presence of non-AGNs in WIBRaLS is estimated to be $< 5\%$ \citep{demenezes2019optical}. The results from our analysis are as follows. 

\begin{itemize}
    \item The total number of associations computed with the LR method is 743 with AP $> 99\%$ , or 1051 associations with AP $> 95\%$. We associate a total of 283 BCUs and 5 UGSs (listed in Tables \ref{tab:BCUs} and \ref{tab:UGSs} in Appendix \ref{ap:tables}). 
    \item Adding a dependence on angular separation to the LR method do not change significantly the results, indicating that WIBRaLS sources are generally in good positional agreement with 4FGL sources. 
    \item A supervised learning method is used for the first time to associate counterparts to $\gamma$-ray sources. As the performance of the algorithm is $\sim 80\%$, the results derived from this method are considered complementary to those obtained with the LR method and should not be interpreted as giving definitive associations to 4FGL sources.
\end{itemize}

This is the first work where spectral properties of blazars are used to associate $\gamma$-ray sources to their low-energy counterparts. Previous methods rely basically on parameters like apparent magnitude and angular separation between the $\gamma$-ray source center and the position of its candidate counterpart.

The lists of BCUs and UGSs available in Appendix \ref{ap:tables} give an excellent opportunity for testing the LR method presented here. In tables \ref{tab:BCUs} and \ref{tab:UGSs}, the targets are sorted alphabetically and can be prioritized by choosing those with higher APs. Optical spectroscopy follow ups of these targets may be crucial to test if the association based on MIR colors is indeed an appropriate approach.

\acknowledgments

This work was supported by FAPESP (Funda\c{c}\~ao de Amparo \`a Pesquisa do Estado de S\~ao Paulo) under grants 2016/25484-9, 2018/24801-6 (R.M.) and 2017/01461-2 (R.N.). The work of F.M. is partially supported by the ``Departments of Excellence 2018-2022'' Grant awarded by the Italian Ministry of Education, University and Research (MIUR) (L. 232/2016) and made use of resources provided by the Compagnia di San Paolo for the grant awarded on the BLENV project (S1618\_L1\_MASF\_01) and by the Ministry of Education, Universities and Research for the grant MASF\_FFABR\_17\_01. F.M. also acknowledges financial contribution from the agreement ASI-INAF n.2017-14-H.0 while A.P. the financial support from the Consorzio Interuniversitario per la Fisica Spaziale (CIFS) under the agreement related to the grant MASF\_CONTR\_FIN\_18\_02. R.D'A. is supported by NASA contract NAS8-03060 (Chandra X-ray Center). This investigation is supported by the National Aeronautics and Space Administration (NASA) grants GO6-17081X and GO9-20083X. We thank the anonymous referee for the constructive comments allowing us to improve the manuscript.

In this publication we made use of data products from the Wide-field Infrared Survey Explorer, which is a joint project of the University of California, Los Angeles, and 
the Jet Propulsion Laboratory/California Institute of Technology, 
funded by the National Aeronautics and Space Administration. We also used \texttt{TOPCAT}\footnote{\url{http://www.star.bris.ac.uk/~mbt/topcat/}} \citep{taylor2005topcat} and \texttt{astropy}\footnote{\url{https://www.astropy.org/index.html}} \citep{astropy2013A&A...558A..33A,astropy2018AJ....156..123A} for preparation and manipulation of the data.

\appendix

\section{List of targets for optical spectroscopic follow up}

\label{ap:tables}

A list with all BCUs associated with the LR method is provided in Table \ref{tab:BCUs}, where the sources are promising targets for future optical spectroscopic follow ups. The columns are the name of the source in $\gamma$-rays and MIR, the AP obtained with the LR method developed here, the AP obtained with the RF approach, the AP obtained with the Bayesian method used in 4FGL and the AP computed with the LR method based on angular distance as in 4FGL. Similarly, Table \ref{tab:UGSs} lists the 5 UGSs described in \S \ref{Results_LR} and \S \ref{Results_RF} and their APs.

\begin{table*}
    \centering
\begin{tabular}{|c|c|r|r|r|r|}
\hline
  \multicolumn{1}{|c|}{4FGL name} &
  \multicolumn{1}{c|}{WISE name} &
  \multicolumn{1}{c|}{LR WISE} &
  \multicolumn{1}{c|}{RF} &
  \multicolumn{1}{c|}{Bayesian} &
  \multicolumn{1}{c|}{LR 4FGL} \\
\hline
    J0001.6-4156 & J000132.74-415525.2 & 0.99 & 0.97 & 1.0 & 0.85\\
    J0008.0-3937 & J000809.17-394522.8 & 0.87 & 0.61 & 0.92 & 0.88\\
    J0010.8-2154 & J001053.64-215704.2 & 0.96 & 0.42 & 0.99 & 0.95\\
     ... & ...  & ... & ...  & ...  & ... \\
    J1238.1-4541 & J123806.03-454129.6** & 0.99 & 1.0 & 1.0 & 0.94\\
    J1239.4+0728 & J123924.58+073017.2* & 1.0 & 0.96 & 1.0 & 0.94\\
     ... &  ... &  ... & ...  & ...  & ... \\
\hline\end{tabular}
    \caption{BCUs associated by the LR method based only on spectral parameters. The BCUs listed here are good targets for future optical spectroscopic follow ups. In the first two columns we have the names of the $\gamma$-ray source and its WISE counterpart. The last four columns show the APs as given by the LR and RF methods in this work and by the Bayesian and LR methods used in 4FGL. Sources tagged with ``*'' and ``**'' are confirmed blazars, with optical spectra available in \protect\cite{deMenezes2020_OptCampX} and Pe\~{n}a-Herazo et al. (2020, in prep), respectively. The full version of this table is available in the online material.}
    \label{tab:BCUs}
\end{table*}


\begin{table*}
    \centering
    \begin{tabular}{|l|l|r|r|}
\hline
  \multicolumn{1}{|c|}{4FGL Name} &
  \multicolumn{1}{c|}{WISE name} &
  \multicolumn{1}{c|}{LR WISE} &
  \multicolumn{1}{c|}{RF} \\
\hline
  J0202.4+2943 & J020239.70+294325.0 & 0.99 & 0.75 \\
  J0620.7-5034 & J062046.13-503350.9 & 0.97 & 0.68 \\
  J1028.1-3550 & J102751.56-355210.7 & 0.98 & 0.31 \\
  J1309.3+8035 & J130628.16+803142.3 & 0.93 & 0.61 \\
  J2013.9-8717 & J201440.97-871707.8 & 0.98 & 0.90 \\
\hline\end{tabular}
    \caption{UGSs associated by the LR method. These sources are good targets for future optical spectroscopic follow ups. In the first two columns we have the names of the $\gamma$-ray source and its WISE counterpart. The last two columns show the AP as given by the LR and RF methods.}
    \label{tab:UGSs}
\end{table*}

\bibliography{sample63}{}
\bibliographystyle{aasjournal}



\end{document}